
\font\ninerm=cmr9
\font\ninei=cmmi9
\font\ninesy=cmsy9
\catcode`@=11
\newskip\ttglue
\def\twelvepoint{\def\rm{\fam0\twelverm}%
  \textfont0=\twelverm \scriptfont0=\ninerm \scriptscriptfont0=\sevenrm
  \textfont1=\twelvei  \scriptfont1=\ninei  \scriptscriptfont1=\seveni
  \textfont2=\twelvesy \scriptfont2=\ninesy \scriptscriptfont2=\sevensy
  \textfont\itfam=\twelveit  \def\it{\fam\itfam\twelveit}%
  \textfont\slfam=\twelvesl  \def\sl{\fam\slfam\twelvesl}%
  \textfont\ttfam=\twelvett  \def\tt{\fam\ttfam\twelvett}%
  \textfont\bffam=\twelvebf  \scriptfont\bffam=\tenbf
   \scriptscriptfont\bffam=\sevenbf  \def\bf{\fam\bffam\twelvebf}%
  \tt \ttglue=.5em plus.25em minus.15em
  \normalbaselineskip=14pt
  \setbox\strutbox=\hbox{\vrule height10pt depth5pt width0pt}%
  \let\sc=\tenrm  \let\big=\twelvebig  \normalbaselines\rm}
\def\twelvebig#1{{\hbox{$\left#1\vbox to 10pt{}\right.\n@space$}}}
\catcode`@=12
\twelvepoint
\font\tenbm=msbm10
\font\sevenbm=msbm7
\newfam\msyfam
\textfont\msyfam=\tenbm
\scriptfont\msyfam=\sevenbm
\def\Bbb#1{{\fam\msyfam #1}}
\def\R{{\Bbb R}}
\def\C{{\Bbb C}}

\def\eth{{\Bbb g}}
\baselineskip 22pt
\def\vsk{\vskip 11pt}
\def\hl{{\hat l}}
\def\hm{{\hat m}}

\def\hn{{\hat n}}
\def\scri{{\cal I}^+{}}

\newcount\seccount
\seccount=0
\newcount\subseccount
\subseccount=0
\newcount\EEK
\EEK=0
\def\eek{\global\advance\EEK by 1\eqno(\the\EEK )}
\def\sec#1{\global\subseccount=0\global\advance\seccount by
1\vsk\vsk\noindent\bf{\uppercase\expandafter{\romannumeral%
\seccount}}.  #1\rm\vsk}
\def\subsec#1{\global\advance\subseccount by
1\vsk\noindent\bf%
\ifcase\subseccount\or A\or B\or C\or D\or E\or F\or G\or H\or I\or J\or
K\or L\or M\or N\fi .  #1\rm\vsk}

\newcount\rtn
\rtn=0
\def\ref{\xdef\ruff{\spacefactor=\the\spacefactor}{\global\advance\rtn by 1}%
${}^{\the\rtn}$\ruff{ }}
\def\nref{\xdef\ruff{\spacefactor=\the\spacefactor}{\global\advance\rtn by 1}%
${}^{\the\rtn}$\ruff}
\def\drtn{\xdef\ruff{\spacefactor=\the\spacefactor}{\global\advance\rtn by 1}%
${}^{\the\rtn{\global\advance\rtn by 1},\,
\the\rtn}$\ruff{ }}
\def\rnn{{\global\advance\rtn by 1}\the\rtn}
\def\fnote#1{{\global\advance\ftn by 1}\baselineskip
 11pt\relax\footnote{${}^{\the\ftn}$}{#1}\baselineskip 22pt{}}
\hbox{ }
\vskip .5in
\centerline{\bf How to Estimate Energy Lost to Gravitational Waves}
\vskip .5in
\centerline{Adam D. Helfer}
\vskip .25in
\centerline{Department of Mathematics, University of Missouri, Columbia, MO
65211}
\vskip .5in
\centerline{\it Abstract\rm}
\vskip .25in

The energy--momentum radiated in gravitational waves by an isolated source is
given by a formula of Bondi.  This formula is highly non--local:  the
energy--momentum is not given as the integral of a well--defined local
density.  It has therefore been unclear whether the Bondi formula can be used
to
get information from gravity--wave measurements.  In this note, we obtain, from
local knowledge of the radiation field, a lower bound on the Bondi flux.


\vfill\eject
\sec{Introduction}

\noindent The theoretical possibility of doing gravity--wave astronomy has been
open since Einstein's prediction of gravitational radiation.  It now seems that
there may be unambiguous detections of gravitational waves in the next
decade.\drtn  This detection would of itself be important --- Einstein regarded
it as a key test of general relativity --- but also we should like to be able
to
extract information about the source of the waves and the space--time through
which they have passed.  For astrophysical sources, it is particularly
desirable
to have model--independent methods of doing this.  In this paper, I shall show
how to get a lower bound on the luminosity of the source in terms of observed
data.  Only very weak general assumptions are made:  that we are very far away
from the source, and that the source does not have a singularity in unobserved
directions.

There are now very good grounds for believing that the correct measure of the
energy carried in gravitational waves is given by a formula of
Bondi.\ref\xdef\Bondiref{\the\rtn}%
It would seem impossible, however, to get a direct measurement of the Bondi
energy
for any astrophysical source.  The main difficulty is that, for its definition,
the Bondi energy requires a knowledge of the asymptotic gravitational field in
all directions.  The problem is not merely that one does not know what the
emitted energy per unit solid angle is in unobserved directions, but that \it
knowledge of the field in these unobserved directions is necessary to properly
\rm define \it the emitted energy per solid angle in the observed directions.
\rm  In other words, the Bondi energy does not arise from a well--defined
energy
density.  Here we show one can get a lower bound on the contribution of local
terms to the total energy.

It should be mentioned that there have been various proposed approximate
local definitions of the energy flux from a source.  Not all of these are
applicable in the asymptotic regime (the weakness of the field may cause
degeneracies in the formalisms).  It seems there are no known approximations
which are even qualitatively controlled in this regime, i.e., for which it is
known whether the approximation provides an under-- or an overestimate.

Since the arguments of this paper depend on an understanding of the
non--locality of the Bondi energy, it is worthwhile to review how this comes
about.  Section II summarizes the evidence that Bondi's formula is the correct
measure of the energy of an isolated system at a retarded time.  Section III
reviews the structure of asymptotically simple space--times, with emphasis on
the places non--local (in angle) considerations enter.  Section IV contains the
estimate of the local contribution to the Bondi flux; the reader already
familiar with the Bondi formalism may skip directly to this.

\it Conventions.  \rm  Our conventions are those of Penrose \&
Rindler,\ref\xdef\PWref{\the\rtn}%
which is also a good reference for the material on the Bondi--Sachs formalism
used here.  The metric has signature $+---$, the curvature tensors satisfy
$[\nabla _a,\nabla _b]v^c=R_{abd}{}^cv^d$, $R_{ac}=R_{abc}{}^b$, $R=R_a{}^a$.
The speed of light is taken as unity.

\sec{Energy--Momentum in General Relativity}

\noindent Because there are no isometries of a general space--time, the usual
special--relativistic theory of energy and momentum as conserved quantities
does
not exist in general relativity.  It is therefore a non--trivial question to
determine what the physically correct measures of these quantities are.
Indeed,
precisely because the usual foundations of the theory are absent, we may wonder
whether there will be any single measure of these quantities which will enjoy
the universal importance energy--momentum in special relativity has.

There are however now very good reasons to believe that the correct measure of
the energy--momentum of an isolated system in exact general relativity at a
give
retarded time (the precise meanings of ``isolated system'' and ``retarded
time''
will be given later) is given by the formula of Bondi:

\itemitem{} (a) It is covariant and it gives physically reasonable answers in
al
known examples.

\itemitem{} (b) There are now three independent derivations of the Bondi
energy--momentum.  (The original by Bondi;${}^\Bondiref
$ the derivation by ``linkages''
of Winicour, Tamburino and Geroch;\drtn  and the derivation by twistor theory
by
Penrose.\nref )

\itemitem{} (c) The theorems of Ludvigsen \& Vickers,\ref  Schoen \& Yau\ref
and others show that the Bondi energy is positive under physically realistic
assumptions, and can become zero only if the space--time is flat in the future.

\itemitem{} (d) A combination of the Arnowitt--Deser--Misner theory\drtn  and
th
analyses of Ashtekar \& Magnon--Ashtekar\ref  and Ashtekar \& Streubel\ref
provides a link with a Hamiltonian theory of conserved quantities for an
asymptotic group of translations.

\noindent It cannot be said, however, that a knowledge of the Bondi
energy--momentum provides us with a full understanding of energy--momentum in
general relativity.  Still unsolved are the important problems of understanding
the energy--momentum at finite distances from sources, and of understanding how
the energy--momenta of subsystems combine.  These are crucial for getting
bounds on the exchange of energy--momentum in astrophysical processes.

\sec{Asymptotically Simple Space--Times}

\noindent In order to model an isolated system in general relativity, one
considers a space--time with certain asymptotics.  For
radiation problems, one requires suitable behavior as one goes outwards in
null directions.  This was first investigated by Bondi and
coworkers${}^{\Bondiref ,\, \rnn ,\,\rnn}$, and is now generally
expressed in the ``scri'' formalism of Penrose.\ref   One requires that
space--time should admit a boundary at future null infinity.

Let $(\widehat M , {\widehat g}_{ab})$ be a space--time.  (The use of hats here
is to avoid a proliferation of them in what follows.)  We say
$(\widehat M , {\widehat g}_{ab})$ is \it future
asymptotically simple \rm if $\widehat{M}$ embeds as the interior of a
manifold--with--boundary $M$, where

(a) There exists a smooth function $\Omega$ positive on $\widehat M$ and zero
on
$\scri =\partial M$, with $\nabla _a\Omega$ not zero on $\scri$, and
such that the conformally rescaled metric $g_{ab}=\Omega ^2{\widehat g}_{ab}$
is
smooth on $M$;

(b) Every null geodesic acquires a future end--point on $\scri$.

\noindent The symbol $\scri$ is read ``scri--plus'' (for ``script I--plus'').
The function $\Omega$ is not determined uniquely by these
requirements; any further rescaling of the form $\Omega\mapsto\omega\Omega$,
where $\omega$ is positive on $\scri$, is allowed.

This definition is in certain respects too simplistic (see Penrose \&
Rindler${}^\PWref$ for criticisms and also for the precise differentiability
requirements), but will be adequate for the purposes of this paper.  We shall
also require that

(c) The stress--energy tensor falls off suitably as one tends to $\scri$
and Einstein's equations hold; and

(d) $\scri$ has topology $S^2\times\R$.

\noindent These follow from certain assumptions on the space--time, but it is
reasonable to take the point of view that these are physical hypotheses for a
well--defined radiation problem.  It should be emphasized that we are not
assuming that physical space--time is future asymptotically simple, but that
isolated sources in general relativity are modeled by future asymptotically
simple space--times.  We are thus working in the approximation in which the
distance of the source to external strong gravitational fields is infinite.
However, there are no restrictions on the internal strengths of the field,
the relative strengths of different multipole moments, the speeds of sources,
etc.

In what follows, much of the analysis will be done in terms of
the rescaled metric.  Geometric quantities defined in terms of this metric
will generally be represented by symbols without hats.  So $\nabla _a$ and
$R_{abcd}$ are the covariant derivative and Riemann tensor associated
with $g_{ab}$, for example; whereas ${\widehat\nabla}_a$ and ${\widehat
R}_{abcd}$ would be the correxponding quantities for the physical metric.

\subsec{Consequences of Future Asymptotic Simplicity}

\noindent We shall now examine the consequences of the hypothesis of future
asymptotic simplicity, and on the way see how the non--locality of
gravitational energy arises.  Here we only outline the results; for details of
the arguments, see Penrose \& Rindler.${}^\PWref$

One can show that (assuming the cosmological constant is zero\ref) the
hypersurface $\scri$ is
null, with the ``$\R$'' factors ruled by null geodesics.  Then
$n^a=-g^{ab}{\nabla}_b\Omega$ is a null
future--pointing tangent to these geodesics.  We coordinatize $\scri$ as $\{
(u,\theta ,\phi )\mid u\in\R\, ,\ (\theta ,\phi )\in S^2\}$, where $u$
satisfies
$n^a{\nabla}_au=1$. (The coordinatization will be fixed more precisely
later.)

A section $u=Z(\theta ,\phi )$ is called a \it cut \rm of $\scri$.  A cut is
the intersection of $\scri$ with an outgoing null hypersurface, and can be
thought of as the asymptotic limit of a wave--front.  Each cut possesses a
conformal structure (that is, the restriction of $g_{ab}$, determined
up to rescalings) and an orientation.  It is well--known that on a
two--dimensional manifold, a choice of conformal structure and orientation is
equivalent to a choice of complex structure.  Thus the cuts of $\scri$ are
naturally equipped with complex structures, and are therefore naturally Riemann
surfaces.  Since they are topologically spheres, they must be
complex--analytically identical to the Riemann sphere.  The group of
structure--preserving transformations of the Riemann sphere is $SL(2,\C )$
modulo $\{\pm 1\}$, and this group is isomorphic to the Lorentz group.  The
Bondi energy--momentum transforms as a four--vector under the action of this
Lorentz group.

(Two comments are in order here.  First, this action of the Lorentz group on a
sphere is quite familiar:  the Lorentz group acts on the celestial sphere in
the same way.  Second, the Einstein equations imply that the field $n^a$ is
shear--free, and this in turn implies that flowing along $n^a$ identifies
the conformal structures on any two cuts.  This allows the Bondi
momentum at one cut to be compared with the Bondi momentum at another cut.)

The argument just given uses the global structure in an essential way.  Suppose
we restricted our attention to a small (connected, simply--connected) patch on
a cut.  In this case, the Riemann Mapping Theorem states that any one such
patch is complex--analytically equivalent to any other such patch.  Thus there
is an infinite--dimensional freedom in embedding such a patch in the Riemann
sphere.  It is impossible to tell which embedding is correct without a
knowledge of the whole sphere.

It is convenient to use the residual freedom in the
conformal factor to arrange that the metrics on the $u=$constant cuts are
those of unit spheres.  We can also arrange that $n^a$ is $g_{ab}$--covariantly
constant over $\scri$.  Then we say we have a \it Bondi coordinate system.  \rm
We form a null tetrad by introducing a future--pointing null vector $l^a$
orthogonal to the $u=$constant cuts and normalized so that
$g_{ab}n^al^b=1$, and also introduce a complex null vector $m^a$
tangent to the cuts, orthogonal (therefore) to both $n^a$ and $l^a$, and
normalized so that $g_{ab}m^a{\overline m}^b=-1$.  (There is also a certain
orientation required for $m^a$; see Penrose \& Rindler.${}^\PWref$)

It is convenient to define an associated tetrad in $M$ for the metric
${\widehat g}_{ab}$:  take ${\hl}^a=\Omega ^2l^a$, ${\hm}^a=\Omega m^a$,
${\hn}^a=n^a$.

\subsec{The Peeling Property}

\noindent The Weyl curvature tensor ${\widehat C}_{abcd}$ is the conformally
covariant part of the Riemann curvature: $C_{abcd}=\Omega ^2{\widehat
C}_{abcd}$; and we have ${\widehat R}_{abcd}={\widehat C}_{abcd}$ in vacuum.
It
is convenient to gather the ten real independent components of the Weyl tensor
into five complex quantities,
$$\eqalign{{\Psi}_0&={C}_{abcd}{l}^am^bl^cm^d\cr
           {\Psi}_1&={C}_{abcd}{l}^am^bl^cn^d\cr
           {\Psi}_2&={C}_{abcd}{l}^am^b{\overline m}^cn^d\cr
           {\Psi}_3&={C}_{abcd}{l}^an^b{\overline m}^cn^d\cr
           {\Psi}_4&={C}_{abcd}{\overline m}^an^b{\overline m}^cn^d\cr
}\eek$$
Then
$${\Psi}_n=\Omega ^{-4+n}{\widehat\Psi }_n\, .\eek$$
(The combinations appear much simpler in spinor notation.)  The Einstein
equations turn out to imply that
$${\Psi}_1={\Psi}_2={\Psi}_3={\Psi}_4=0 \hbox{ on }\scri\, .\eek$$
The assumption that $\scri$ has topology $S^2\times\R$ then turns out to force
${\Psi}_0$ to vanish at $\scri$.  Then in fact it can be shown that the
quantities
$$\psi _n=\Omega ^{-1}{\Psi }_n=\Omega ^{-5+n}{\widehat\Psi}_n\eek$$
are smooth on $M$.  It can also be shown that $\Omega$ can be taken to
go like $1/r$, where $r$ is an affine parameter along a geodesic with tangent
$l^a$.  Therefore the physical quantities have the asymoptotic behaviors
$${\widehat\Psi}_n=O(r^{-5+n})\, .\eek$$
This is the \it Sachs peeling property.  \rm  The \it radiation field \rm is
${\widehat\Psi}_4$; we expect it to be dominant at large distances.

In this argument, non--local considerations entered in an essential way.
We had to appeal to the topology of the ``$S^2$'' factor in order to deduce the
required asymptotic behavior of the Weyl tensor.  Thus if the source is
isolated in the sense that it is future asymptotically simple, then its Weyl
tensor will have certain asymptotics.  Although this is an interesting
mathematical deduction, it is not (as phrased) too significant physically,
since it is reasonable to take the position that we are interested only in
sources which are isolated and whose Weyl tensors fall off appropriately
anyway.  On the other hand, the contrapositive has some significance.  If we
observed a source for which the peeling property did not hold in some
direction, then we could conclude that it could not be asymptotically simple:
it would necessarily  be non--isolated in some perhaps unobserved
direction.\ref

\subsec{The Shear and the News}

\noindent The fields $\psi _n$ are related on $\scri$ by the Bianchi
identities.  For our purposes, it is enough to note that in a Bondi frame one
has
$$\psi _3=\eth N\hbox{ and }\psi _4=\partial N/\partial u\, ,\eek$$
where $N$ is quantity called the \it Bondi news function.  \rm  (Here $\eth$ is
a derivative in the $m^a$ direction.)  The news may be regarded as a certain
component of the curvature tensor of the metric $g_{ab}$.  This however
does not give much insight into its significance.  A more direct interpretation
is that it measures (minus) the rate of change of the (conjugate of the) shear
of the outgoing null hypersurfaces meeting $\scri$ in $u=$constant cuts,
$$N=-\partial{\overline\sigma} /\partial u\, .\eek$$

One can of course transform all these quantities back to arbitrary, non--Bondi,
coordinate systems.  Then the Bondi news is still defined as a certain
component of the curvature, but it plays a less direct role in the analysis.
The rate of change of the shear (which is no longer equal to this curvature
component) is the more important quantity.  Although the shear itself \it is
\rm conformally covariant, this time derivative of it is not.  The problem is
precisely that the choice of conformal factor $\omega$ determines which cut is
to be regarded as being to the future one unit of time from a given cut.

It is instructive to examine the rate of change of (minus the conjugate of) the
shear in an arbitrary, non--Bondi system.  It turns out to be
$$\omega ^{-2}\bigl[ {\overline N}-\eth ^2\log\omega
+(\eth\log\omega )^2\bigr]\eek$$
when the conformal factor is $\omega \Omega$.
The quantity
$$-\eth ^2\log\omega +(\eth\log\omega )^2\eek$$
vanishes precisely when $\omega$ has the form induced by a Lorentz motion.
Thus under Lorentz transformations, which take one Bondi system to another, the
news transforms as a con\-form\-ally--weighted quantity.  However, under
general
changes, it is altered in a more complicated, non--linear, way.

\subsec{The Bondi Momentum}

\noindent The space of asymptotic translations can be identified with
quantities of the form $\xi n^a$, where $\xi$ is a linear combination of the
$l=0$ and $l=1$ spherical harmonics.  The compontent of the Bondi
four--momentum
in the $\xi n^a$ direction at a given cut is
$$P_\xi =(8\pi G)^{-1}\int\bigl[ \psi _2-\sigma N\bigr] \xi\, d{\cal S}$$
and $d{\cal S}=\sin\theta d\theta d\phi$.  For sources outside of the solar
system, we expect not to be able to evaluate this, since according to the
peeling property, we expect ${\widehat\Psi}_2$ to fall off like $1/r^3$.
However, it can be shown that
$$\partial _uP_\xi =(8\pi G)^{-1}\int |N|^2\xi\, d{\cal S}\, .\eek$$
In particular, the power (for which we take $\xi =1$) is
$$(8\pi G)^{-1}\int |N|^2\, d{\cal S}\eek$$\xdef\bndpwr{\the\EEK}%
in the frame of the Bondi system.

\sec{If Only the Radiation Field is Known}

\noindent Let us assume an isolated gravitational source, which may be
represented mathematically by a future asymptotically simple space--time, as
above.  On account of the peeling property, which gives the asymptotic behavior
$${\widehat\Psi}_n=O( r^{n-5})\hbox{ as }r\to\infty\, ,\eek$$
we expect that for all sources outside the solar system only ${\widehat\Psi}_4$
will be measurable in the near future.  In this case, the only constraint on
the news is the equation
$$\psi _4=\partial  N/\partial u\, .\eek$$\xdef\eee{\the\EEK}%
Suppose for the moment that $\psi _4$ is known on some patch of $\scri$ of
small angular extent $U$ and for some interval $u_0\leq u\leq u_1$ of retarded
time.  Then the news function can be recovered up to the ambiguity in
integrating (\eee ).  Let us set
$$N_0(u)=\int _{u_0}^u\psi _4(u)\, du\, .\eek$$
Then
$$N=N_0+\alpha\, ,\eek$$
where $\alpha$ is a function of the angular variables only.  The contribution
to the emitted energy from $U$ is
$$(8\pi G)^{-1}\int _{u_0}^{u_1} \int _U |N_0+\alpha |^2\, d{\cal S}du\,
.\eek$$
This is a quadratic form in $\alpha$, which is bounded below (by zero).
By elementary calculus of variations, we find it has a stationary point (which
must be a minimum) at
$$\alpha =-(u_1-u_0)^{-1}\int _{u_0}^{u_1}N_0\, du\, .\eek$$
The minimum value is
$$\Delta E=(8\pi G)^{-1}\int _{u_0}^{u_1}|N_0|^2\, d{\cal S}du
  -(8\pi G)^{-1}(u_1-u_0)^{-1}\int \bigl| \int _{u_0}^{u_1}N_0du\bigr| ^2
  \, d{\cal S}\, .\eek$$
In terms of the physical, unrescaled, field, we have
$$\eqalign{N_0&=\int _{u_0}^u\Omega ^{-1}{\widehat\Psi}_4(u)\, du\cr
  &=r\nu (u)\cr}\eek$$
where
$$\nu (u)=\int _{u_0}^u\Psi _4(u)\, du\, .\eek$$
Then the emitted energy per unit area is at least
$${{d\Delta E}\over{dA}}_{\rm min}=(8\pi G)^{-1}\int _{u_0}^{u_1} |\nu |^2 du
  -(8\pi G)^{-1}(u_1-u_0)^{-1}
  \left| \int _{u_0}^{u_1}\nu du\right| ^2\, .\eek$$
This is our main result.

It should be emphasized that this contribution to the emitted energy refers to
the frame defined by a Bondi system compatible with the laboratory frame, and
no
to the frame defined by the Bondi four--momentum of the source.  The latter
would be more desirable for most theoretical purposes, but cannot be found
from local data.  However, we can get a bound involving this quantity and the
relative boost of the frames.  Let $\Delta P_a=P_a(u_1)-P_a(u_0)$.  If
$\Delta E_B$ is the time--component of this in the Bondi frame, and $\beta$ is
the relative speed of the two frames, then we have
$$\sqrt{\Delta P_a\Delta P^a}\cosh\beta =\Delta E_B\, .\eek$$
Since we have a lower bound on $\Delta E_B$, we get a lower bound on
$\sqrt{\Delta P_a\Delta P^a}\cosh\beta$.

Currently--planned gravitational--wave detectors operate by measuring
geo\-de\-sic deviation.  Let $(e_t^a,e_x^a,e_y^a,e_z^a)$ be an orthonormal
tetra
transported parallel along the center of the detector.  Let us choose the
orientation so that the source is on the negative $z$--axis.
Then the components of
the curvature which are detected are
$e_t^ae_t^cR_{abcd}$.  In our case, only $\Psi _4$ contributes to this, and we
have
$$e_t^ae_t^cR_{abcd} =({\rm Re}\, \Psi _4)E^+_{bc}
  +({\rm Im}\, \Psi _4)E^\times _{bd}\, ,\eek$$
where
$$\eqalign{E^+_{bd}&=e^x_be^x_d-e^y_be^y_d\cr
  E^\times _{bd}&=e^x_be^y_d+e^y_be^x_d\cr}\eek$$
are the ``plus'' and ``cross'' polarizations.  Let us consider
detection of a wave at a single frequency, for example.  We take
$$\Psi _4=a\cos (\omega u)\, ,\eek$$
and find
$$
{{d\Delta E}\over{dA}}_{\rm min} =(16\pi G)^{-1}|a|^2\Delta u
\bigl[ 1 +T\bigr]\, ,\eek$$
where $\Delta u=u_1-u_0$ and the last term,
$$\eqalign{T&=(2\omega\Delta u)^{-1}
[\sin (2\omega u_1)-\sin (2\omega u_0)]\cr &\qquad
  -(\omega\Delta u)^{-2}[2-\cos (2\omega u_1)-\cos (2\omega u_0)
       -4\sin (\omega u_1)\sin (\omega u_0)]\, ,\cr}\eek$$
represents oscillatory terms.
{}From this, we see that the ``plus'' and ``cross'' modes contribute equally
and
without interference to the lower bound on the energy emitted (for
monochromatic
waves), as well as that we must have $\omega\Delta u{\lower
.5ex\hbox{${\buildre
>\over\sim}$}}1$ to get an estimate where the oscillatory contributions are
small.

\vsk
\it Acknowledgement.  \rm  I thank the referee for constructive criticism.

\vfill\eject\frenchspacing

\parshape=0
\global\parindent=0pt
\newcount\arno
\arno=0
\everypar{{\global\advance\arno by
1}\hangafter=1\hangindent=1pc{}${}^{\the\arno}$}

A. Abramovici et al., Science \bf 256\rm , 325 (1992).

K. S. Thorne, in \it Recent Advances in Relativity, \rm eds. A.~I.~Janis and
J.~R.~Porter (Birkh\"auser, Boston, Basel, Berlin, 1992).

H. Bondi, Nature \bf 186\rm , 535 (1960).




R. Penrose and W. Rindler, \it Spinors and Space--Time,
\rm (Cambridge, University Press, 1984--6).

J. Winicour and L. Tamburino, Phys. Rev. Lett. \bf 15\rm , 601 (1965).

R. Geroch and J. Winicour, J. Math. Phys. \bf 22\rm , 803 (1981).

R. Penrose, Proc. R. Soc. Lond. \bf A381\rm , 53 (1982).

M. Ludvigsen and J. A. G. Vickers, J. Phys. \bf A14\rm , L389 (1981).

R. Schoen and S. T. Yau, Phys. Rev. Lett. \bf 48\rm , 369 (1982).

R. Arnowitt, S. Deser and C. W. Misner, J. Math. Phys. \bf 1\rm , 434 (1960).

T. Regge and C. Teitelboim, Ann. Phys. \bf 88\rm , 286 (1974).

A. Ashtekar and A. Magnon--Ashtekar, Phys. Rev. Lett. \bf 43\rm , 181 (1979).

A. Ashtekar and M. Streubel, Proc. R. Soc. Lond. \bf A376\rm , 585 (1981).



H. Bondi, M. G. J. van der Burg and A. W. K. Metzner, Proc. R. Soc.
Lond. \bf A269\rm , 21 (1962).

R. K. Sachs, Proc. R. Soc. Lond. \bf A270\rm , 103 (1962).

R. Penrose, in \it Relativity,
Groups and Topology: \rm the 1963 Les Houches Lectures, eds. B.~S.~DeWitt and
C.~M.~DeWitt.  (Gordon and Breach, New York, 1964).

Strictly speaking, the hypothesis is not that the cosmological constant
is zero for the whole space--time, but that it is zero for this model of an
isolated system.  If there were a cosmological constant for the whole
space--time, one would expect the present approximation to apply to sources
which were far from us on the scale of their characteristic dimensions, but
close on the scale of the radius of curvature set by the cosmological constant.

Note however that verification of the peeling property requires
knowledge of the field at arbitrarily great distances from the source.

\bye